\documentclass[a4paper,amsmath,preprintnumbers,showpacs,twocolumn,prl,superscriptaddress, nofootinbib]{revtex4}
\usepackage{amssymb}
\usepackage{graphicx}
\usepackage{epstopdf}
\usepackage{bm}
\usepackage{epsfig}
\usepackage{graphics}
\usepackage{xspace}
\usepackage{amsfonts}
\usepackage{verbatim}
\usepackage{natbib}
\usepackage{color}

\newcommand{\nn}{\nonumber}
\newcommand{\cut}{\textrm{cut}}
\newcommand{\mO}{\mathcal{O}}

\begin{document}


\title{Fully differential Higgs pair production in association with a $Z$ Boson at next-to-next-to-leading order in QCD}

\author{Hai Tao Li}\email{haitaoli@lanl.gov}
\affiliation{
Los Alamos National Laboratory, Theoretical Division, Los Alamos, NM 87545, USA}
\affiliation{ARC Centre of Excellence for Particle Physics at the Terascale, School of Physics and Astronomy, Monash University, Victoria, 3800 Australia}
\author{Chong Sheng Li}\email{csli@pku.edu.cn}
\affiliation{
Department of Physics and State Key Laboratory of Nuclear Physics and Technology, Peking University, Beijing 100871, China}
\affiliation{Center for High Energy Physics, Peking University, Beijing 100871, China}
\author{Jian Wang}\email{j.wang@tum.de}
\affiliation{Physik Department T31,  Technische Universit\"at M\"unchen, James-Franck-Stra\ss e~1,
D--85748 Garching, Germany }



\begin{abstract}
We present a fully differential next-to-next-to-leading order QCD calculation of the Higgs pair production in association with a $Z$ boson
at hadron colliders, which is important for probing the trilinear Higgs self-coupling.
The next-to-next-to-leading-order corrections enhance the next-to-leading order  total cross sections by a factor of $1.2\sim 1.5$, depending on the collider energy, and change the shape of next-to-leading order  kinematic distributions.
We  discuss how to  determine  the trilinear Higgs self-coupling using our results.

\end{abstract}

\pacs{12.38.Bx,14.80.Bn}

\maketitle

\section{Introduction}
The discovery of the Higgs boson at the LHC \cite{Aad:2012tfa,Chatrchyan:2012ufa} is a milestone 
toward  understanding the mechanism of electroweak symmetry breaking. 
The accumulated data indicate that it is a spin-$0$ and $CP$-even particle with a mass of 125 GeV \cite{Khachatryan:2014kca}.
The couplings of this particle with massive vector bosons and fermions have been measured to agree with the standard model (SM) expectations \cite{Aad:2013wqa,CMS-PAS-HIG-14-009}.
The next step is to measure these couplings  more precisely and to probe its self-couplings,
which is mandatory to clarify the Higgs potential, and thus electroweak symmetry breaking mechanism.
This is one of the main tasks of the LHC \cite{Djouadi:1999rca,Baur:2002rb,Baur:2002qd,Baur:2003gp,Dolan:2012rv,Papaefstathiou:2012qe,Baglio:2012np,Barger:2013jfa,Barr:2013tda,Dolan:2013rja,Englert:2014uqa,
Liu:2014rva,deLima:2014dta,ATL-PHYS-PUB-2014-019,CMS-PAS-FTR-15-002}
and also a future 100 TeV hadron collider
\cite{Yao:2013ika,Barr:2014sga,Li:2015yia,Azatov:2015oxa,Papaefstathiou:2015iba,Zhao:2016tai}.

The trilinear Higgs self-coupling can be measured in two ways.
One is the indirect detection from its effect on the single Higgs prductions and decays 
\cite{McCullough:2013rea,Castilla-Valdez:2015sng,Gorbahn:2016uoy,Degrassi:2016wml,Bizon:2016wgr,DiVita:2017eyz,Maltoni:2017ims}
or the electroweak precision observables \cite{Degrassi:2017ucl,Kribs:2017znd}.
The other is the direct measurement of the Higgs pair productions at high-energy colliders 
\cite{Djouadi:1999rca,Baur:2002rb,Baur:2002qd,Baur:2003gp,Dolan:2012rv,Papaefstathiou:2012qe,Baglio:2012np,Barger:2013jfa,Dolan:2013rja,Barr:2013tda,Yao:2013ika,Frederix:2014hta,deLima:2014dta,Englert:2014uqa,Liu:2014rva,Barr:2014sga,Li:2015yia,Azatov:2015oxa,Papaefstathiou:2015iba,Zhao:2016tai,Cao:2015oaa,Cao:2015oxx,Cao:2016zob}.
The dominant production channel at a hadron collider is the gluon-gluon fusion $gg\to hh$ process which involves a top-quark loop.
The other channels, including the vector boson fusion, the vector boson associated production and the top quark pair associated production,
have relatively smaller cross sections. 
However, the additional particles in the final state make it easier to distinguish the signal and background events.  Actually,  the different channels have different characteristics, and thus are complementary to each other and deserve discussion on the same footing. 

\begin{figure}
  \includegraphics[width=0.9\linewidth]{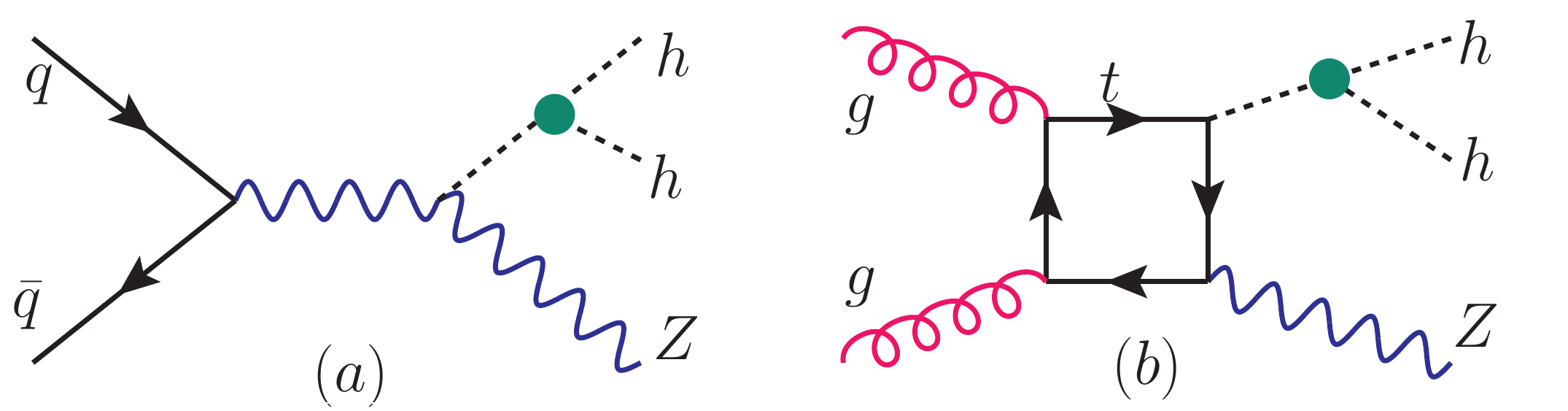}\\
  \caption{Sample Feynman diagrams for  $pp\to Zhh$ production. }
  \label{fig:top_loop}
\end{figure}

The vector boson associated production channel, as shown in Fig.~\ref{fig:top_loop},
is of special interest for several reasons.
First, the final-state vector boson provides a characteristic tag of the event 
so that large quantum chromodynamics (QCD) backgrounds can be suppressed.
As a result, one can select the events with  the Higgs boson pair decaying to $b\bar{b}b\bar{b}$, 
which has the largest branching ratio,
and thus the cross section of this channel is comparable to that of $gg\to hh$
with  the Higgs boson pair decaying to $\gamma\gamma b\bar{b} $ \cite{Cao:2015oxx}.
Second, all the involved  Higgs couplings in this channel are not loop-induced up to next-to-leading order (NLO)
(which is the case in the gluon-gluon fusion channel),
avoiding the unknown  effects of virtual heavy particles.
It is more direct and clear to interpret the cross section as a function of the Higgs couplings.
Finally, it depends on the value of the Higgs self-coupling in a different form, compared to
the gluon-gluon fusion $gg\to hh$ channel.
In particular, it is  sensitive to the Higgs self-coupling which is larger than the SM value \cite{Frederix:2014hta,Cao:2015oxx}.

Precise theoretical predictions  of the vector boson associated Higgs pair productions are crucial for a proper interpretation of experimental data.
The total cross sections  have been calculated up to next-to-next-to-leading order (NNLO) in QCD  \cite{Baglio:2012np}.
However, after experimental cuts are imposed on the final state,
it is not clear whether the NNLO QCD corrections are the same over the full phase space.
Therefore it is desirable to provide fully differential NNLO QCD calculations.
We have presented NNLO differential cross sections of the
Higgs pair production in association with a $W$ boson at hadron colliders
in a previous work \cite{Li:2016nrr}.
In this Letter, we report the first fully differential NNLO QCD calculation  of the Higgs pair production in association with
 a $Z$ boson,
which is important for probing the trilinear Higgs self-coupling at the LHC or future high-energy hadron colliders.
In contrast to the $pp\to Whh$ process, the cross section of $pp\to Zhh$ receives a large contribution from the 
loop-induced process $gg\to Zhh$ starting from NNLO, as shown in Fig.~\ref{fig:top_loop}(b).
This additional contribution has a significant impact  on the total and differential cross sections.
Our  calculation  shows that the NNLO corrections indeed change the shape of NLO kinematic distributions.
In the peak region of some differential distributions, the NNLO corrections reach up to $80\%$, compared to NLO results.
Therefore, our result is an important ingredient for extracting information on the Higgs self-couplings.

\section{The method}
We perform the (N)NLO QCD differential calculations  using the  $q_T$ subtraction method  \cite{Catani:2007vq}, where $q_T$ denotes the transverse momentum of the final-state colorless system.
This method is based on the understanding of the cross section near the infrared divergent  regions,
i.e.,  $q_T\to 0$.
When $q_T$ is small,  the cross section can be factorized as a combination of  a hard function, a soft function and  transverse-position dependent parton distribution functions (PDFs). The soft function and transverse-position dependent  PDFs describe the low-energy dynamics near the infrared divergent region, which  are  independent of the high-energy  dynamics except for the dependence due to collinear anomaly \cite{Becher:2010tm}, and thus can be considered as universal. 
Most of them have been obtained up to NNLO, based on which a large number of processes
have been calculated at (N)NLO \cite{Catani:2007vq,Catani:2009sm,Catani:2011qz,Cascioli:2014yka,Ferrera:2014lca,Gehrmann:2014fva,Campbell:2016yrh,Grazzini:2016swo,Grazzini:2016ctr,deFlorian:2016uhr}.

In our calculation, we divide the (N)NLO cross section into two parts by a cutoff parameter $q_T^{\cut}$.
In one part with $q_T<q_T^{\cut}$,
the cross section can be obtained by expanding the transverse momentum resummation formula up to NNLO
\footnote{Notice that the $gg\to Zhh$ channel does not depend on $q_T^{\cut}$
and is computed apart from the $q_T$ subtraction method. As a result, 
we neglect this channel in all description and discussion about the $q_T$ subtraction method,
but include its contribution in the numerical results. }.
The other part  with $q_T>q_T^{\cut}$ is just the NLO correction to $pp\to Zhhj$,
which can be calculated using standard NLO subtraction method, 
such as the FKS \cite{Frixione:1995ms} or the dipole subtraction~\cite{Catani:1996vz}.

Let us first discuss the part with  $q_T<q_T^{\cut}$.
We make use of the transverse momentum resummation
based on  soft-collinear effective theory (SCET)~\cite{Bauer:2000yr,Bauer:2001yt,Beneke:2002ph}.
Since the process of $q\bar{q}\to Zhh$ can be considered as a production of an off-shell $Z^*$ boson and its decay to $Zhh$,
the cross section of $q\bar{q}\to Zhh$ production in the small $q_T$ region
can be written as~\cite{Becher:2010tm}
\begin{multline}
  \frac{d\sigma_{Zhh}}{dq_T^2 dydM^2} =\frac{1}{2 SM^2} \sum_{i,j=q,\bar{q},g}\int_{\zeta_1}^1 \frac{dz_1}{z_1}
      \int_{\zeta_2}^1 \frac{dz_2}{z_2}\int d\Phi_3
     \\
      \times
     \bigg( f_{i/N_1}(\zeta_1/z_1,\mu)f_{j/N_2}(\zeta_2/z_2,\mu) H_{q\bar{q}}(M,\mu)
      \\ 
      \times  C_{q\bar{q}\leftarrow i j}(z_1,z_2,q_T, M,\mu)\bigg) \left[1
+\mathcal{O}\left(\frac{q_T^2}{M^2}\right) \right]~,
\label{eq:dsigma}
\end{multline}
where $q_T$, $M$ and $y$ are the transverse momentum, invariant mass and rapidity of the $Zhh$ system,
$ d\Phi_3$ the three-body phase space,
and $ f_{i/N}(x,\mu)$  the standard PDF.
The integration lower limits $\zeta_1$ and $\zeta_2$ are determined by $S,M,q_T, y$.
The hard function $H_{q\bar{q}}(M,\mu) $ contains the contribution from hard-scale interactions and
is extracted by matching the (axial) vector currents in QCD onto effective currents built out of fields in SCET.
The two-loop results can be obtained from the hard function of a Drell-Yan process~\cite{Becher:2006mr}.
All the $q_T$-dependent terms are contained in the collinear kernel~\cite{Becher:2010tm}
\begin{align} \label{eq:c_scet}
    & C_{q\bar{q}\leftarrow i j}(z_1,z_2,q_T, M,\mu)
     = \frac{1}{4\pi}\int d^2x_\perp e^{- i x_\perp \cdot q_\perp}  \\
    &  \left(\frac{x_T^2 M^2}{b_0^2}\right)^{F_{q\bar{q}}(x_T^2, \mu)}
     I_{q\leftarrow i}(z_1,L_\perp, \alpha_s) I_{\bar{q}\leftarrow j}(z_2,L_\perp, \alpha_s)\nn
\end{align}
with  $x_T^2=-x_{\perp}^2$, $  b_0 = 2 e^{-\gamma_E}$ and $  L_\perp = \ln\frac{x_T^2\mu^2}{b_0^2}$.
The  function $F_{q\bar{q}}(x_T^2, \mu)$ arises from the effect of collinear anomaly~\citep{Becher:2010tm}.
The function $I_{q\leftarrow i}$ describes the evolution of a parton $i$ to $q$ at fixed $x_{\perp}$,
of which the two-loop results  can be found in Refs.~\cite{Gehrmann:2012ze,Gehrmann:2014yya}~.
With all the NNLO ingredients available  it is straight forward to perform the integration of $q_T$ from $0$ to $q_T^{\cut}$ in Eq.(\ref{eq:dsigma}).

 Next, we turn to the other part of the cross section with large $q_T$.
 Due to the constraint  $q_T>q_T^{\rm cut}$, there must be at least one parton in the final state.
 The (N)LO cross section of $pp\to Zhhj$ contributes to the (N)NLO cross section of
 $pp\to Zhh$. 
Therefore, we need to calculate only the NLO corrections to $pp\to Zhhj$ production.
This is one of the advantages by using $q_T$ subtraction, i.e.,
the present tools to perform NLO calculations can be utilized without any substantial change.
Notice that it is even unnecessary to apply any jet algorithms in the final state and all the infrared singularities are either regularized by the cut $q_T>q_T^{\rm cut}$ or cancelled between the virtual and real corrections,
since the constraint $q_T> q_T^{\cut}$ prevents the momentum of the harder parton  in the final state 
from being arbitrarily soft or collinear to the initial-state momenta. 
The combination of phase spaces of $pp \to Zhhj$ at NLO with large $q_T$ and
$pp \to Zhh$ at NNLO with small $q_T$ recovers the whole phase space of $pp\to Zhh$ at NNLO.
In this work, we use the modified MadGraph5\_aMC@NLO \cite{Alwall:2014hca} to 
calculate the (N)LO corrections to $pp \to Zhhj$.

\begin{figure*}
  \includegraphics[width=0.4\linewidth]{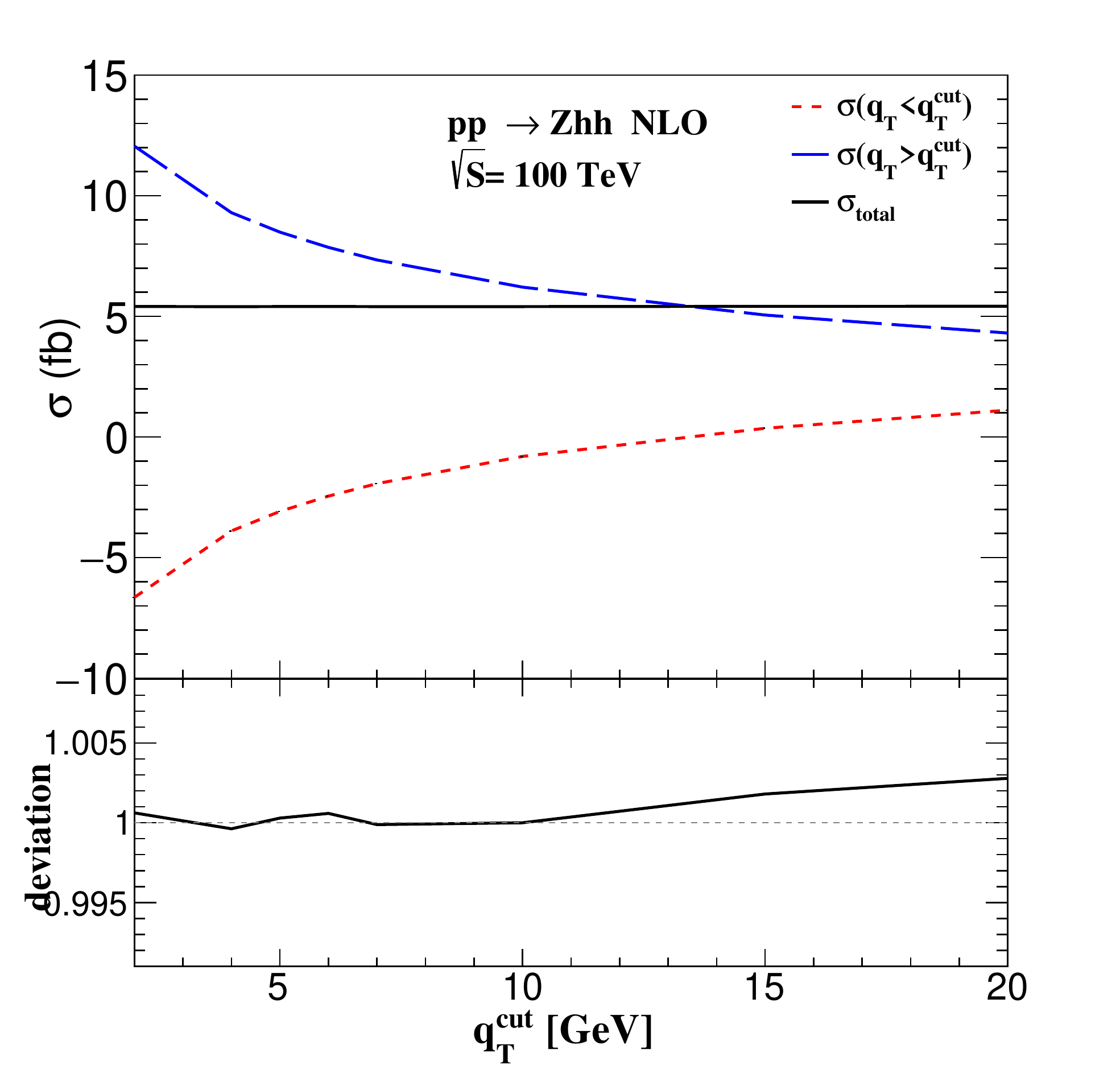}
    \includegraphics[width=0.4\linewidth]{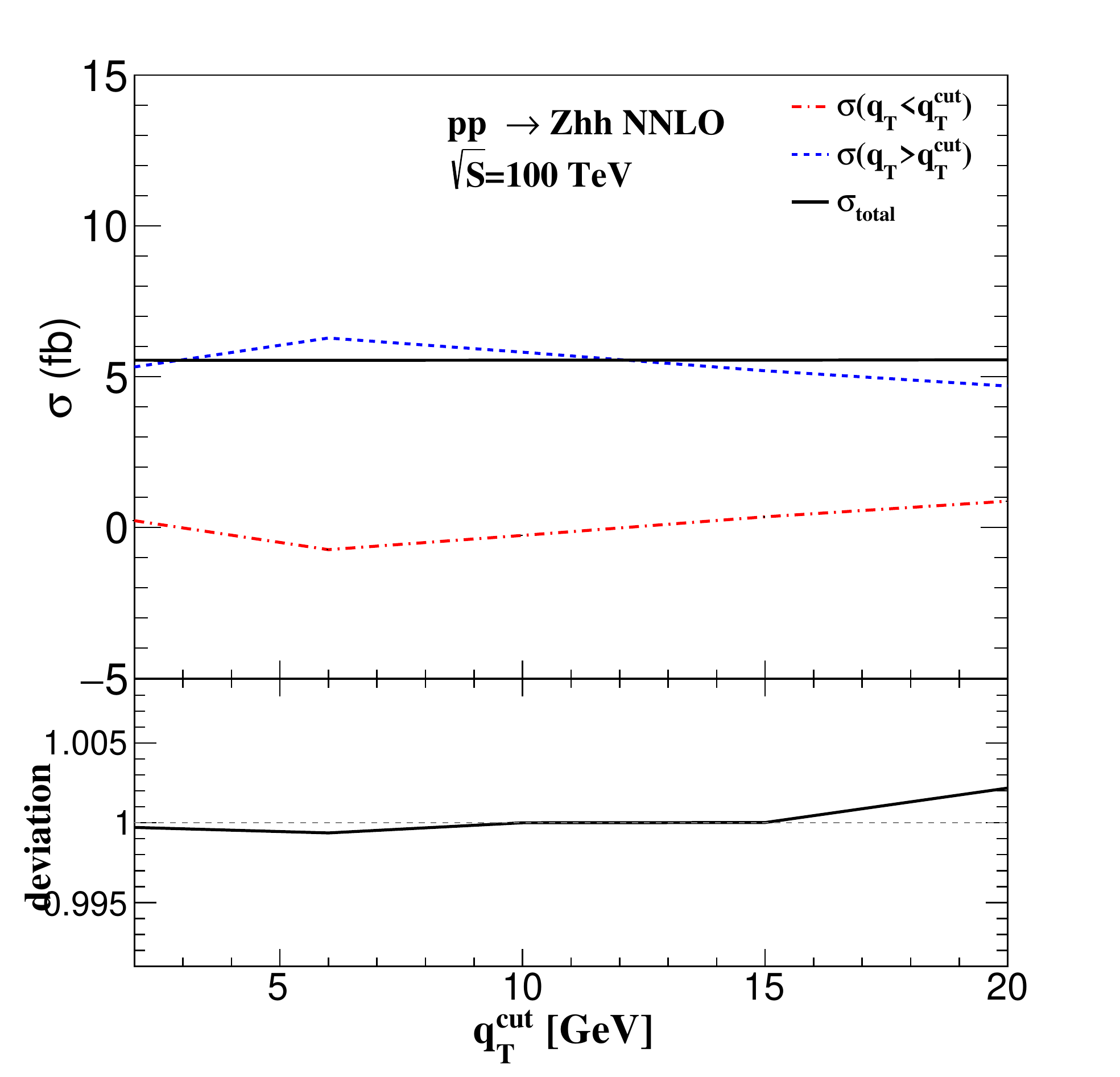}\\
  \caption{The total cross sections of $pp\to Zhh$ production at NLO (left) and NNLO (right) without contribution from loop-induced  $gg$ fusion channel.
    In the bottom panels, the deviation is defined as $\sigma(q_T^{\rm cut})/\sigma(q_T^{\rm cut}=10~{\rm GeV})$.  The curves are drawn  using the linear interpolation method.}
  \label{fig:qTcut}
\end{figure*}

Combining the two parts together, we obtain the (N)NLO  QCD differential cross section of the process $pp\to Zhh$
\begin{multline}\label{eq:main}
   \frac{d\sigma_{Zhh}}{d\Phi_3dy}\Big\vert_{\textrm{(N)NLO}} =
   \\
  \int_0^{q_T^{\cut}} dq_T \frac{d\sigma_{Zhh}^{\rm (N)NLO}}{d\Phi_3 dy dq_T} 
   + \int^{q_T^{\max}}_{q_T^{\cut}} dq_T \frac{d\sigma_{Zhhj}^{\rm (N)LO}}{d\Phi_3dydq_T}~,
\end{multline}
where $q_T^{\max}$ is fixed by the partonic center-of-mass energy and the phase space constraints.
The cross section of the loop-induced process $gg\to Zhh$   is both ultraviolet and infrared finite,
and thus there is no need to introduce a new subtraction method to calculate this process. 
We also use MadGraph5\_aMC@NLO \cite{Alwall:2014hca,Hirschi:2015iia} to compute this contribution.
We have compared our results with Ref.~\cite{Agrawal:2017cbs} and found agreement within uncertainties.

\begin{figure}
    \includegraphics[width=0.9\linewidth]{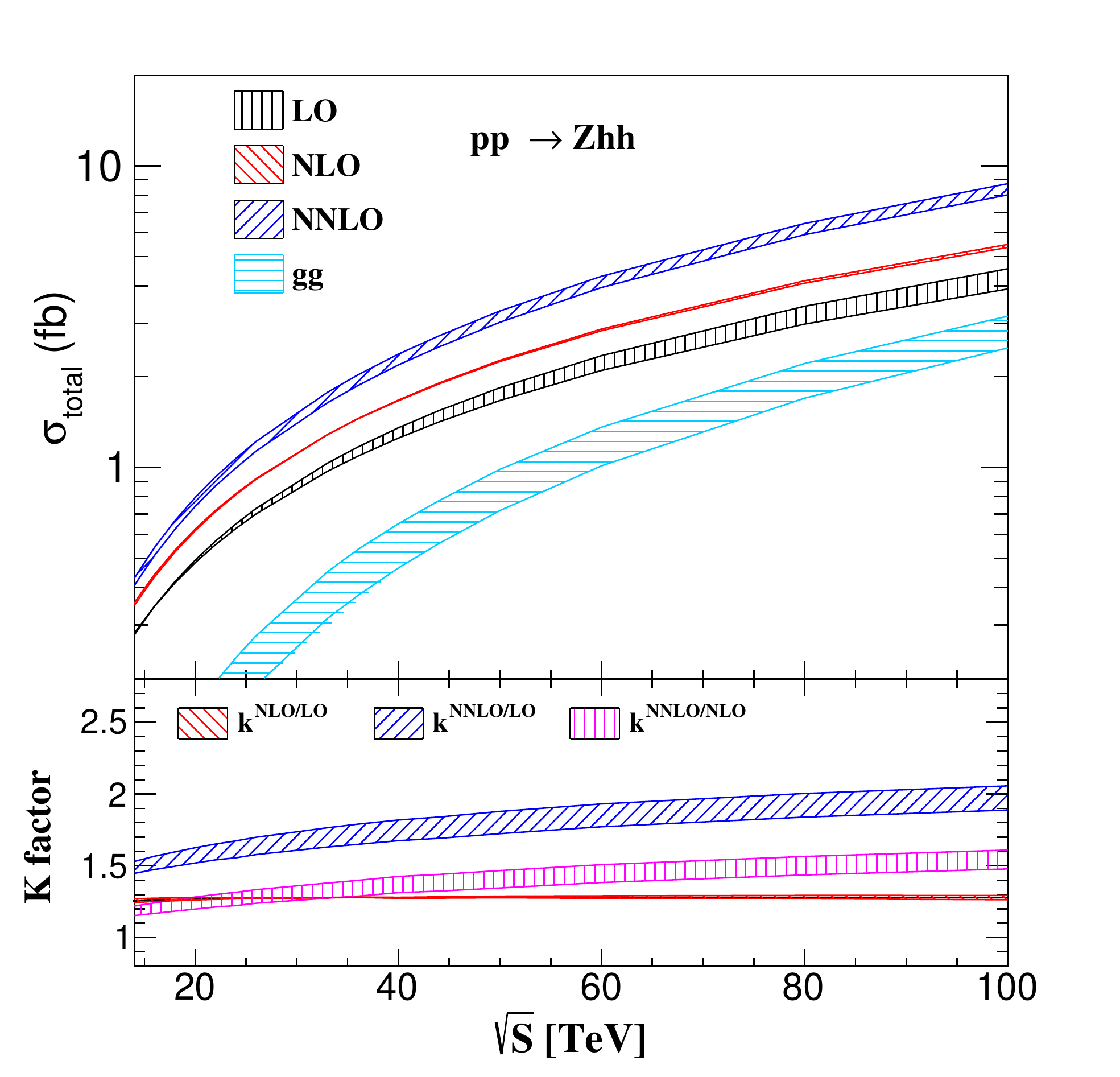}\\
  \caption{The total cross section of $pp\to Zhh$ production as a function of the collision energy.
The bands denote the scale uncertainties when varying $\mu=\mu_F=\mu_R$ by a factor of two.  The NNLO total cross sections include the loop-induced $gg$ channel,  and the contribution from this channel is also shown in the upper panel individually.  }
  \label{fig:sqrts}
\end{figure}

\section{Numerical results}
In the following of this paper, we present  numerical results for $Zhh$ production at the proton-proton colliders 
with $\sqrt{S}=$ 14 TeV and 100 TeV. The  CT14 PDF sets~\cite{Dulat:2015mca} and the associated strong coupling are used 
throughout our calculation. 
In particular, we use the LO, NLO, and NNLO PDF sets for the corresponding LO, NLO, and NNLO calculations of the cross section,
respectively.
The default factorization scale $\mu_F$ and renormalization scale $\mu_R$ are both set equal to $M$  to avoid possible large logarithms. 
The scale uncertainties are estimated by varying the default value by a factor of two up and down.
The SM parameters are given by
\begin{align}
 & M_Z = 91.1876 ~\textrm{GeV}, ~
 m_h = 125 ~\textrm{GeV},~
 m_t = 173.5 ~\textrm{GeV},
 \nn \\
& \sin^2 \theta_W = 0.222,~~ \alpha = \frac{1}{132.5},  ~~ \lambda_{hhh}^{\rm SM} = \frac{m_h^2}{2v},
\end{align}
where $v=246$ GeV is the vacuum expectation value of the Higgs field. 

When using the $q_T$ subtraction method  in Eq.~(\ref{eq:main}), 
the first priority is to check that the total cross section is independent of the cut-off parameter $q_{T}^{\cut}$. 
Figure~\ref{fig:qTcut} shows the predictions from the two parts with $q_T<q_{T}^{\cut}$ and $q_T>q_{T}^{\cut}$
individually, as well as their sum, at both NLO and NNLO. 
Here, the loop-induced  $gg$ fusion channel is not included in the NNLO result because it is  independent of $q_T^{\cut}$.
The two parts at NLO 
change dramatically when varying $q_{T}^{\cut}$ from 2 GeV to 20 GeV, while the NLO total correction is independent of the cut-off parameter. 
At NNLO, the two parts change slightly because the $\mO(\alpha_s^2)$ $q_{T}^{\cut}$-dependent corrections are  almost
equal to the $\mO(\alpha_s)$ ones, but with a  relative minus sign.
 In Fig.~\ref{fig:qTcut} the statistical uncertainties of the total cross section are less than 0.2\%.
Notice that the part with $q_T<q_{T}^{\cut}$ is only accurate at  the leading power of $q_T^{\rm cut}/M$. 
The power corrections are about $(q_T^{\rm cut}/M)^2\sim 0.04\%$ for the choice of $q_T^{\rm cut}=10$~GeV  and a  typical   $M\sim 500$~GeV.
In the following discussion  we choose $q_T^{\rm cut}=$  10 GeV.
As a cross check, we have compared the NLO  cross section  of $pp\to Zhh$ calculated by Eq.(\ref{eq:main})
with that by the standard NLO program MadGraph5\_aMC@NLO  and found good agreement.

In Fig.~{\ref{fig:sqrts}},  we present the total cross section at different collision energies as well as the $K$-factors of higher-order corrections. 
With the increasing of collision energy,  the cross section increases significantly. 
Compared to the  leading order (LO) results, 
the NLO cross sections have much smaller scale uncertainties, especially when the collision energy is large. 
The NLO $K$-factors are around $1.26$ for $14~{\rm TeV}\leq\sqrt{S}\leq 100~{\rm TeV}$. 
The NNLO corrections can enhance the NLO cross section further by a factor of $1.2\sim 1.5$, depending on the collision energy, but have  larger scale uncertainties, about $\pm 5\%$. This is mainly due to the very large contribution from the loop-induced $gg\to Zhh$ process, which is about $13\%$ (14~TeV) $\sim$ $34\%$ (100 TeV) of the NNLO total cross section as shown in Fig.~\ref{fig:sqrts}. In order to reduce the theoretical uncertainty, it is desired to include higher-order corrections to this process, which means that one needs to calculate  two-loop Feynman diagrams of a $2\to 3$ process with three different masses. As far as we know, this is beyond current techniques, and we leave it to future work.
By adopting the same input parameters, 
our calculations of the total cross sections are consistent with those in Ref.\cite{Baglio:2012np}. 

\begin{figure}
    \includegraphics[width=0.9\linewidth]{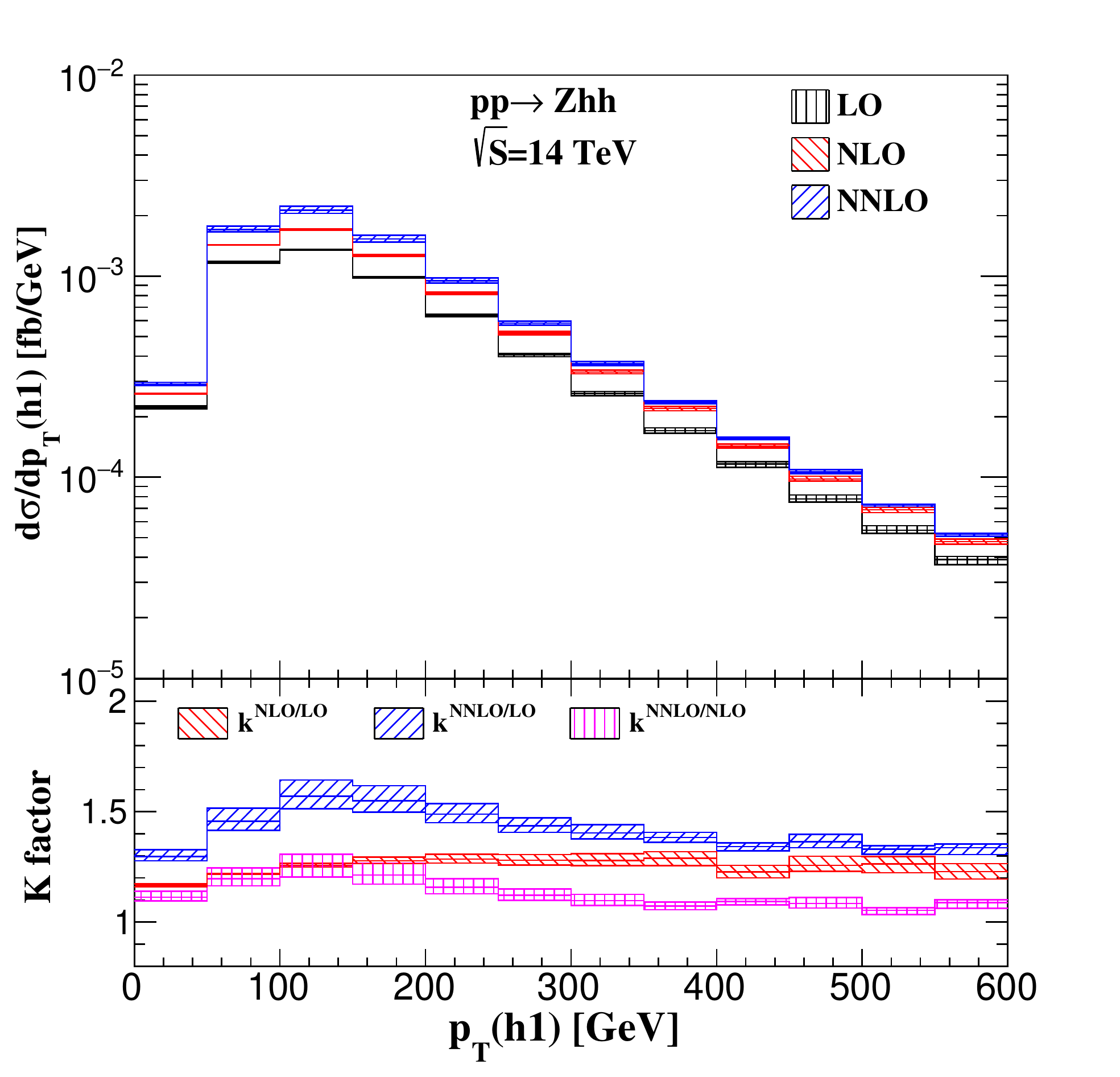}
    \\
  \caption{The kinematic distributions of $pp\to Zhh$ production at the 14 TeV LHC.
  $h1$ denotes the Higgs boson with larger transverse momentum. }
  \label{fig:kin}
\end{figure}

\begin{figure}
    \includegraphics[width=0.95\linewidth]{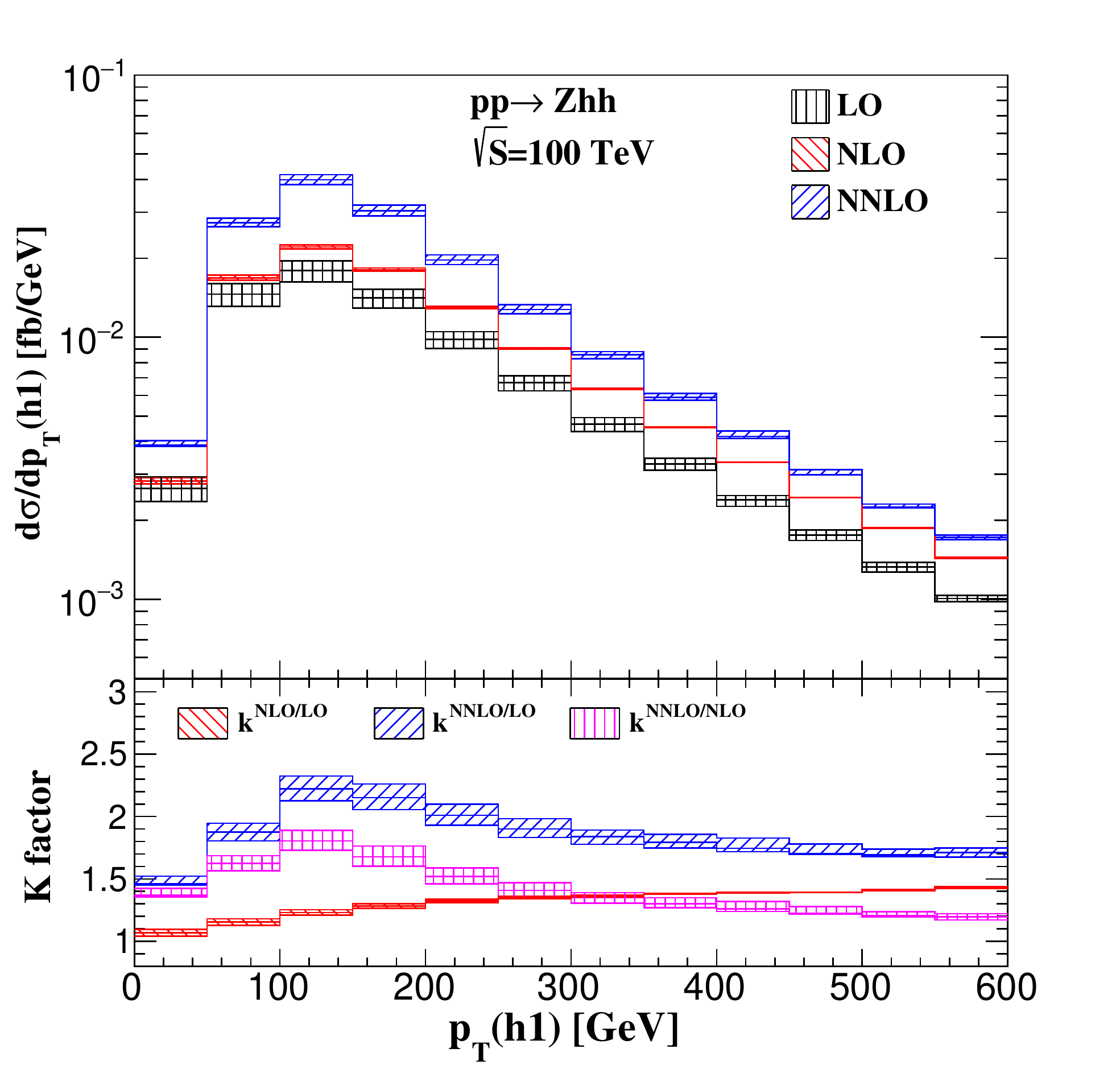}
\\
  \caption{The kinematic distributions of $pp\to Zhh$ production at a future 100 TeV hadron collider.
   $h1$ denotes the Higgs boson with larger transverse momentum.  }
  \label{fig:kin100}
\end{figure}

In Fig.~{\ref{fig:kin}}, we show the transverse momentum $p_T$ distributions of the leading Higgs (the one with larger transverse momentum) at the 14 TeV LHC, which have not be obtained in previous calculations. It can be seen that  the shape of the $p_T$ distribution is nearly unchanged from LO to NLO, 
but at NNLO the peak region increases more significantly than the tail region; i.e., 
the shape of the kinematic distribution is changed. Similarly to the total cross section, the NNLO corrections are very large, and  the scale uncertainties are also  larger than NLO ones because of the contribution from the loop-induced process. Figure~{\ref{fig:kin100}} shows the same distributions at a 100 TeV proton-proton collider. The kinematic features are the same as Fig.~\ref{fig:kin} except for larger NNLO corrections. 
 In particular,
the differential NNLO/NLO $K$-factor in the peak region is as large as 1.8.

\begin{figure}
  \includegraphics[width=0.95\linewidth]{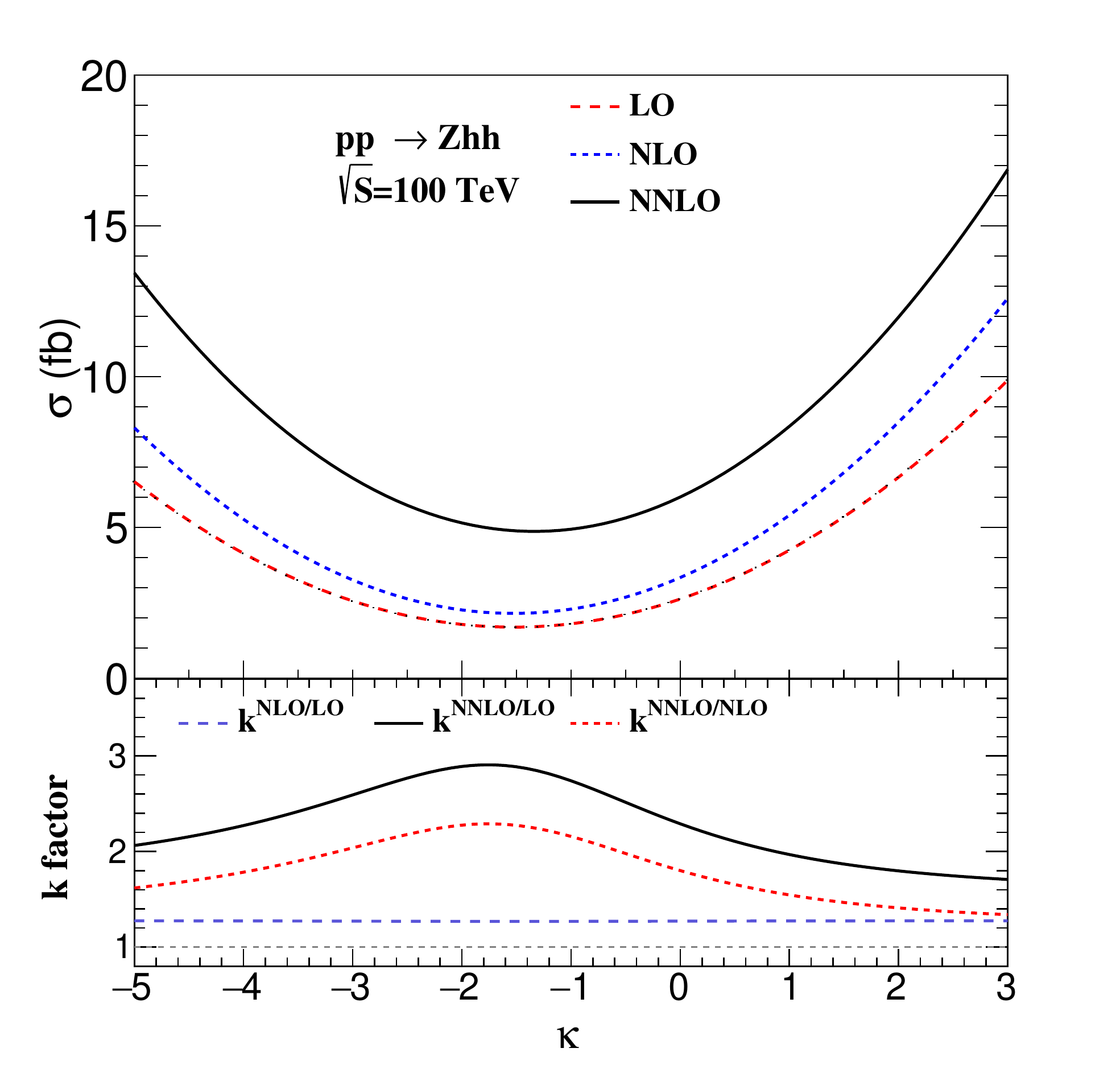}
  \caption{ The total cross section as a function of $\kappa$ at a hadron collider with $\sqrt{S}=100$ TeV. }
  \label{fig:kappa_dep}
\end{figure}

\begin{figure}
  \includegraphics[width=0.95\linewidth]{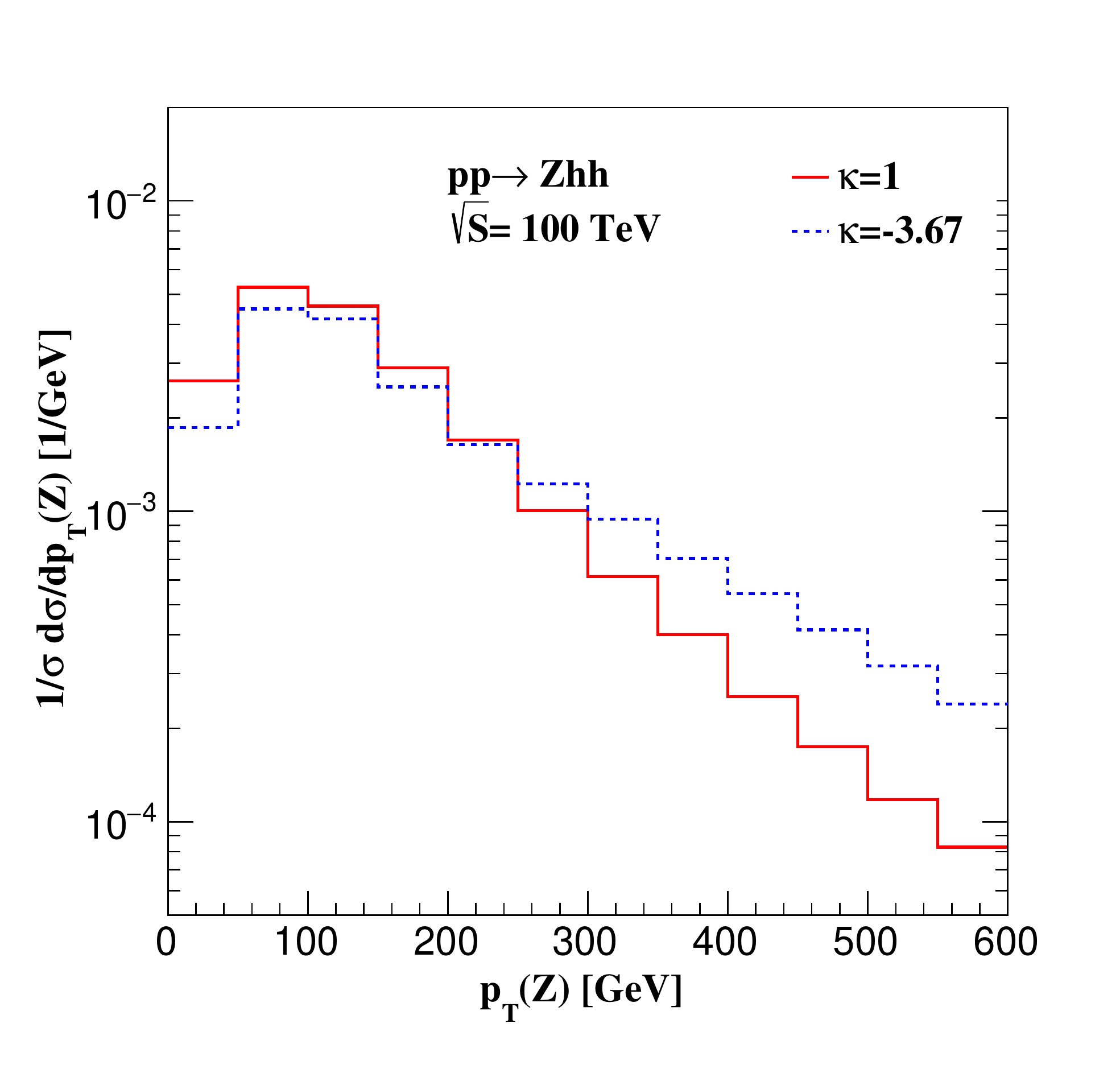}
  \caption{ Normalized NNLO $p_T$ distributions of the $Z$ boson with $\kappa=1$ and $-3.67$, which correspond to the same total cross section. }
  \label{fig:kappa_dep_diff}
\end{figure}

Regarding that a $Z$ boson associated Higgs pair production can be used to probe the trilinear Higgs self-coupling, we investigate the dependence of the total cross section on the self-coupling at a 100 TeV hadron collider in Fig.~\ref{fig:kappa_dep}, where the factor $\kappa$ is defined as
\begin{align}
\lambda_{hhh}=\kappa \lambda_{hhh}^{\rm SM}.
\end{align}
It can be seen  that the total cross section changes by about $-40\%$ to $+100\%$ in the range $-5<\kappa<3$, compared to the SM prediction, 
and thus provides a potential method to measure the trilinear Higgs self-coupling. 
However,   there are, in general, two values of $\kappa$ derived from the same total cross section. 
For example, the NNLO SM total cross section corresponds to $\kappa=1$ and $\kappa=-3.67$.
Therefore, we need other observables, e.g., the differential distributions, to ascertain this self-coupling. Figure~\ref{fig:kappa_dep_diff} shows the normalized NNLO transverse momentum distributions of the $Z$ boson at a 100 TeV collider with  $\kappa=1$ and $\kappa=-3.67$, respectively. 
The SM prediction ($\kappa=1$) has a larger peak but a smaller tail compared with the case of  $\kappa=-3.67$.

\section{Conclusions}
In this paper, we have presented complete NNLO QCD predictions for the total and differential cross sections of $Zhh$ production at hadron colliders  using the transverse momentum  subtraction method.
The NNLO corrections enhance the NLO total cross sections by a factor of $1.2\sim 1.5$, depending on the collider energy, 
and change the shape of NLO kinematic distributions.
We also investigate the sensitivities of the total and differential cross sections to the trilinear Higgs self-coupling,
and find that both of them are needed in order to fix  this self-coupling. 
Our precise predictions will be  helpful for the extraction of information on the Higgs trilinear self-coupling in future experimental analyses.

\emph{Acknowledgements}.---We are grateful to Jun Gao and Hua Xing Zhu for helpful comments on our manuscript. 
Most of our calculations were carried out on the T30 cluster  at the Physics Department of  Technische Universit\"at M\"unchen.
HTL was supported in part by the ARC Centre of Excellence for Particle Physics at the Tera-scale and by the DOE Early Career.
JW was supported by  the BMBF project No. 05H15W0CAA.
This work was also supported in part
by the National Nature Science Foundation of China,
under Grants No. 11375013.

\end{document}